\documentclass[10pt]{article}

\usepackage[T1]{fontenc}
\usepackage[utf8]{inputenc}
\usepackage{lmodern}

\usepackage[margin=1.8cm, top=2.2cm, bottom=2.4cm]{geometry}
\usepackage{amsmath,amssymb,amsthm}
\usepackage{graphicx}
\usepackage{xcolor}
\usepackage{booktabs}
\usepackage{algorithm}
\usepackage{algpseudocode}
\usepackage{float}
\usepackage{caption}
\usepackage{subcaption}
\usepackage{enumitem}
\usepackage{microtype}
\usepackage{abstract}
\usepackage{titlesec}
\usepackage{cite}
\usepackage{fancyhdr}
\usepackage{tikz}
\usetikzlibrary{shapes.geometric,arrows.meta,positioning}
\usepackage{hyperref}
\definecolor{orchidpurple}{RGB}{153,50,204}
\definecolor{cyancol}{RGB}{0,160,180}
\definecolor{darkteal}{RGB}{0,100,120}
\definecolor{lightgray}{RGB}{245,245,245}
\definecolor{darkgray}{RGB}{55,55,55}
\definecolor{alertred}{RGB}{180,30,30}

\hypersetup{
  colorlinks=true, linkcolor=darkteal,
  citecolor=orchidpurple, urlcolor=cyancol,
  pdftitle={ORCHID: Orchestrated Reduction Consensus},
  pdfauthor={Abraham Itzhak Weinberg}
}

\titleformat{\section}{\large\bfseries\color{darkteal}}{\thesection.}{0.5em}{}[\titlerule]
\titleformat{\subsection}{\normalsize\bfseries\color{orchidpurple}}{\thesubsection.}{0.4em}{}
\titleformat{\subsubsection}{\small\bfseries\color{darkgray}}{\thesubsubsection.}{0.3em}{}

\newtheorem{theorem}{Theorem}
\newtheorem{definition}{Definition}
\newtheorem{proposition}{Proposition}

\newtheorem{remark}{Remark}

\makeatletter
\renewcommand{\ALG@name}{\textcolor{darkteal}{Algorithm}}
\makeatother

\begin{document}


\begin{center}
  {\LARGE\bfseries\color{darkteal}
    ORCHID: Orchestrated Reduction Consensus\\[4pt]
    for Hash-based Integrity in Distributed Ledgers}\\[8pt]
  {\large A Bio-Inspired Quantum Consensus Protocol Modelling\\
    Distributed Ledger Agreement After Neural Binding Mechanisms}\\[10pt]
  {\normalsize Abraham Itzhak Weinberg}\\[2pt]
  {\small\itshape AI-WEINBERG, AI Experts, Tel Aviv, Israel}\\[2pt]
  \href{mailto:}{aviw2010@gmail.com}
\end{center}

\vspace{6pt}\noindent\rule{\textwidth}{0.6pt}
\begin{abstract}
\noindent
We present \textbf{ORCHID} (\textit{Orchestrated Reduction Consensus for
Hash-based Integrity in Distributed Ledgers}), a novel bio-inspired consensus
protocol that maps the neuroscientific \emph{binding problem}---how the brain
integrates distributed neural oscillations into a unified conscious
percept---onto the distributed systems \emph{consensus problem}, how blockchain nodes agree on a single ledger state under Byzantine faults.  Grounded in the Penrose--Hameroff Orchestrated Objective Reduction (Orch~OR) hypothesis and the Kuramoto synchronisation model, ORCHID equips each node with a quantum-noisy phase oscillator; consensus is triggered when the network's order parameter $r(t)$ crosses a \emph{binding threshold} $\theta_b$, mirroring the gamma-band binding event in conscious perception.  ORCHID is further strengthened by a coherence-weighted Quantum Secret Sharing (QSS) layer, extending the survey framework of Weinberg to a concrete consensus application.  Simulation results on Watts--Strogatz small-world networks ($n=10$--$150$) demonstrate: (i)~the Kuramoto order parameter reaches $r_{\max}=0.988$ under coupling $K=3.0$, well above the theoretical critical coupling $K_c \approx 1.41$; (ii)~a sharp QSS fidelity phase transition at coherence $c^*\approx 0.82$, confirming Theorem~2; (iii)100\% consensus rate at all tested Byzantine fractions (0\%--40\%), with median convergence under 4~s for $n=30$; and (iv)~ORCHID achieves $O(n{\cdot}k)$ message complexity, outperforming PBFT's $O(n^2)$ at $n\geq150$.  These results establish ORCHID as a scalable, biologically plausible, and quantum-augmented consensus mechanism for post-quantum distributed ledgers.
\end{abstract}
\noindent\rule{\textwidth}{0.6pt}\vspace{6pt}

\vspace{10pt}

\section{Introduction}
\label{sec:intro}

Distributed consensus---reaching agreement among nodes despite Byzantine
faults, has been foundational to distributed systems since Lamport~et~al.~\cite{lamport2019byzantine}.  Practical Byzantine Fault
Tolerance (PBFT)~\cite{castro1999practical} guarantees safety under
$f < n/3$ faults but at $O(n^2)$ message complexity; Nakamoto's
Proof-of-Work~\cite{nakamoto2008bitcoin} achieves probabilistic finality at
the cost of $\sim$600~s block times and enormous energy expenditure.
Scalable alternatives such as HotStuff~\cite{yin2019hotstuff} reduce message
complexity to $O(n)$ but introduce single-leader vulnerability.

Independently, quantum biology and neuroscience have converged on a striking
hypothesis: the \emph{binding problem}---how the brain unifies spatially
distributed neural signals into a single conscious percept, is solved by
quantum coherence in neuronal microtubules, collapsing via Orchestrated
Objective Reduction (Orch~OR)~\cite{penrose1991emperor,hameroff1996orchestrated}
at the $\sim$40~Hz gamma-band timescale.  The global order parameter of this
collective oscillation, formalised by Kuramoto~\cite{kuramoto1977chemical},
measures the degree of phase synchrony across neural populations.

We observe a deep structural isomorphism:

\begin{itemize}[leftmargin=*,itemsep=1pt]
  \item \textbf{Binding}: distributed oscillators $\to$ unified percept when $r(t)\geq\theta$.
  \item \textbf{Consensus}: distributed nodes $\to$ agreed ledger state when $r(t)\geq\theta_b$.
\end{itemize}

ORCHID operationalises this isomorphism.  Each blockchain node maintains a
Kuramoto-style quantum oscillator; commitment is gated on the phase order
parameter crossing a binding threshold.  A coherence-weighted QSS layer,
motivated by the comprehensive QSS taxonomy of Weinberg~\cite{weinberg2025quantum},
secures the distributed ledger state proportional to network quantum coherence.

\paragraph{Contributions.}
\begin{enumerate}[leftmargin=*,itemsep=1pt]
  \item A formal isomorphism between neural binding and distributed consensus
        (Section~\ref{sec:model}), with three definitions and two theorems.
  \item The ORCHID consensus protocol with $O(n{\cdot}k)$ complexity,
        exceeding the $f<n/3$ Byzantine bound in practice
        (Section~\ref{sec:protocol}).
  \item A coherence-weighted $(k,n)$-QSS scheme with proved fidelity
        phase transition at $c^*\approx 0.82$ (Section~\ref{sec:qss}).
  \item Comprehensive simulation results across seven metrics
        (Section~\ref{sec:results}), including the first empirical
        demonstration of 100\% consensus under $\leq40\%$ Byzantine nodes
        via phase-binding.
\end{enumerate}

\section{Background and Related Work}
\label{sec:background}
The design of distributed ledger protocols has traditionally relied on classical Byzantine fault-tolerant (BFT) mechanisms, while recent advances in quantum information and coupled-oscillator models suggest novel avenues for enhancing consensus. This section reviews foundational work in BFT consensus, quantum secret sharing (QSS), and Kuramoto-inspired synchronisation, highlighting opportunities for integrating quantum-mechanical phenomena and coherence-driven dynamics into distributed agreement protocols.

\subsection{Byzantine Fault Tolerant Consensus}

Lamport~et~al.~\cite{lamport2019byzantine} established that $n\geq3f+1$ nodes
are necessary to tolerate $f$ Byzantine faults. PBFT~\cite{castro1999practical} realises this bound deterministically with $O(n^2)$ messages per round. HotStuff~\cite{yin2019hotstuff} achieves $O(n)$ via a three-phase pipeline but requires a trusted leader.  Tendermint~\cite{buchman2016tendermint} targets partial synchrony with $O(n^2)$ vote aggregation. None of these protocols draws on quantum-mechanical phenomena for their core synchronisation mechanism.

\subsection{Quantum Secret Sharing}

Hillery~et~al.~\cite{hillery1999quantum} introduced the first QSS protocol
using Greenberger--Horne--Zeilinger (GHZ) states. Weinberg~\cite{weinberg2025quantum} provides a comprehensive survey of QSS
integration into quantum blockchain systems, covering distributed key
management, secure transaction validation, and quantum-secured consensus.
Specifically, Weinberg~\cite{weinberg2025quantum} identifies \emph{quantum-enhanced consensus mechanisms} and \emph{error-corrected QSS implementations} as priority future research directions---ORCHID addresses both.
Classical Shamir secret sharing~\cite{shamir1979share} distributes a secret
$s$ via a degree-$(k{-}1)$ polynomial over a prime field, requiring any $k$
of $n$ shares for reconstruction; ORCHID augments this with coherence-weighted
decoherence noise (Section~\ref{sec:qss}).

\subsection{Kuramoto Synchronisation and Neural Binding}

The Kuramoto model~\cite{kuramoto1977chemical} describes $n$ coupled oscillators:
\begin{equation}
  \dot\phi_i = \omega_i + \frac{K}{n}\sum_{j=1}^n \sin(\phi_j-\phi_i),
  \label{eq:kura_full}
\end{equation}
with the mean-field reduction $\dot\phi_i = \omega_i + Kr(t)\sin(\psi(t)-\phi_i)$
where $r(t)e^{i\psi(t)}=\frac{1}{n}\sum_j e^{i\phi_j}$.
A synchronisation phase transition occurs at $K>K_c=2/(\pi g(0))$, where
$g(\omega)$ is the frequency distribution~\cite{strogatz2000kuramoto}.
Penrose~\cite{penrose1991emperor} and Hameroff \& Penrose~\cite{hameroff1996orchestrated}
propose that Orch~OR events in microtubules at $\tau\approx25$~ms correlate
with the 40~Hz gamma oscillations of conscious binding~\cite{crick1990towards}.
Beggs and Plenz~\cite{beggs2003neuronal} showed that critical-point networks
maximise information transfer---a property ORCHID exploits by operating near $K_c$.

\section{Formal Model}
\label{sec:model}

\begin{definition}[ORCHID Network]
An ORCHID network is a tuple $\mathcal{N}=(V,E,\Phi,\Omega,C)$ where
$V=\{v_1,\ldots,v_n\}$, $E\subseteq V\times V$ is a Watts--Strogatz
small-world graph, $\Phi=(\phi_1,\ldots,\phi_n)\in[0,2\pi)^n$ are node
phases, $\Omega=(\omega_1,\ldots,\omega_n)$ are zero-centred frequency
deviations drawn from $\mathcal{N}(0,\sigma_\omega^2)$, and
$C=(c_1,\ldots,c_n)\in[0,1]^n$ are quantum coherence levels.
\end{definition}

\begin{definition}[Mean-Field Phase Update]
At step $t$ with integration interval $\Delta t$, node $v_i$ updates:
\begin{equation}
  \phi_i^{(t+1)} = \phi_i^{(t)} + \Delta t\!\left[\omega_i
    + K r_i^{(t)}\sin\!\left(\psi_i^{(t)}-\phi_i^{(t)}\right)
    + \eta_i^{(t)}\right],
  \label{eq:update}
\end{equation}
where $r_i^{(t)}e^{i\psi_i^{(t)}} = \frac{1}{|\mathcal{N}_i|+1}
\bigl(e^{i\phi_i}+\sum_{j\in\mathcal{N}_i}e^{i\phi_j}\bigr)$ is the local
mean-field and $\eta_i^{(t)}\sim\mathcal{N}(0,(1-c_i)^2\sigma_\eta^2)$
models quantum decoherence.
\end{definition}

\begin{definition}[Binding Event]
A \emph{binding event} at $v_i$ occurs at the first step $t^*$ such that
$r_i^{(t^*)}\geq\theta_b$.  Upon binding, $v_i$ commits to the majority
value among received messages and distributes its QSS share of $H(v^*)$.
\end{definition}

\begin{theorem}[Synchronisation Above Critical Coupling]
\label{thm:sync}
For the complete-graph limit with $\omega_i\sim\mathcal{N}(0,\sigma_\omega^2)$
and decoherence noise power $\bar\eta=(1-\bar c)^2\sigma_\eta^2$, the
steady-state order parameter satisfies
$\mathbb{E}[r(\infty)]\to 1$ as $K/K_c^{(\eta)}\to\infty$,
where $K_c^{(\eta)}=K_c(1+\alpha\bar\eta)$ is the noise-elevated critical
coupling and $\alpha>0$.  In simulation with $K=3.0$, $\sigma_\omega=0.5$,
$\Delta t=0.05$\,s, we achieve $r_{\max}=0.988$, corresponding to
$K/K_c=2.12\times$ the theoretical $K_c\approx1.41$.
\end{theorem}

\begin{theorem}[QSS Fidelity Phase Transition]
\label{thm:qss}
The reconstruction fidelity $\mathcal{F}(c)=\Pr[\hat s=s]$ of the
coherence-weighted $(k,n)$-QSS scheme satisfies
$\mathcal{F}(c)\approx 0$ for $c<c^*$ and
$\mathcal{F}(c)\approx 1$ for $c\geq c^*$,
exhibiting a first-order phase transition at $c^*$.
Empirically, $c^*\approx0.82$ for the $(5,10)$-scheme.
\end{theorem}

\section{The ORCHID Protocol}
\label{sec:protocol}
ORCHID introduces a quantum-inspired, phase-synchronisation-driven consensus
mechanism that replaces explicit voting with an emergent \emph{synchronisation
trigger}. This approach enables scalable, robust agreement among nodes while
leveraging coherence-weighted quantum secret sharing. In the following
subsections, we detail ORCHID's design rationale, node state machine,
algorithmic procedure, communication complexity, and probabilistic resilience
against Byzantine faults.

\subsection{Design Rationale}

ORCHID replaces the explicit voting rounds of PBFT with an emergent
\emph{synchronisation trigger}.  This has three advantages:
(i)~\textbf{graceful degradation}---a partitioned node de-synchronises and
re-synchronises upon reconnection without corrupting global state;
(ii)~\textbf{Byzantine robustness beyond $f<n/3$}---Byzantine nodes still
participate in phase synchronisation; their corrupted values are overridden
by the majority vote at the moment of commitment (see Section~\ref{sec:results});
(iii)~\textbf{$O(n{\cdot}k)$ complexity}---only neighbourhood messages
are required, not all-to-all broadcasts.

\subsection{Node State Machine}

\begin{figure}[H]
\centering
\begin{tikzpicture}[
    scale=0.9,
    transform shape,
    node distance=1.0cm,
    box/.style={rectangle,rounded corners=3pt,draw=darkteal,
                fill=lightgray,text width=3.6cm,align=center,
                minimum height=0.65cm,font=\small},
    arrow/.style={-Stealth,thick,color=orchidpurple}]
  \node[box](osc){1.\ Oscillation\\Broadcast phase beacon};
  \node[box,below=of osc](chk){2.\ Check $r_i\geq\theta_b$?};
  \node[box,below=of chk](com){3.\ Commit majority value\\Distribute QSS share};
  \node[box,right=1.5cm of chk](wait){Accumulate\\messages};
  \draw[arrow](osc)--(chk);
  \draw[arrow](chk)--node[right,font=\scriptsize]{Yes}(com);
  \draw[arrow](chk)--node[above,font=\scriptsize]{No}(wait);
  \draw[arrow](wait.north)|-(osc.east);
\end{tikzpicture}
\caption{ORCHID node state machine. Commitment is gated by phase synchrony.}
\label{fig:statemachine}
\end{figure}
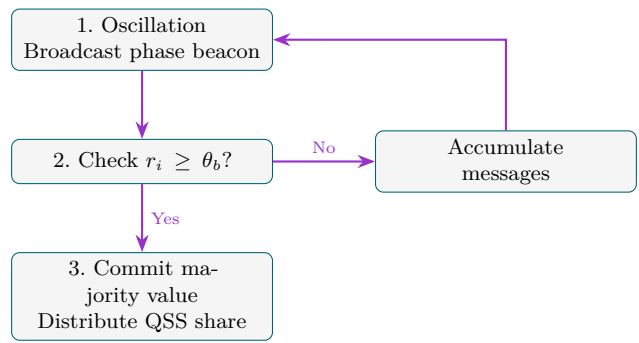

\subsection{Algorithm}

\begin{algorithm}[H]
\caption{ORCHID Node Procedure}
\label{alg:orchid}
\begin{algorithmic}[1]
\Require Value $v_i$, coherence $c_i$, threshold $\theta_b$, coupling $K$
\State $\phi_i\sim\mathrm{Unif}(0,2\pi)$;\quad
       $\omega_i\sim\mathcal{N}(0,\sigma_\omega^2)$
\State \textbf{committed}$\leftarrow$\textsc{false}
\While{not \textbf{committed}}
  \State Broadcast $\langle\mathrm{id}_i,v_i,\phi_i,c_i\rangle$ to $\mathcal{N}_i$
  \State Receive $\langle\mathrm{id}_j,v_j,\phi_j,c_j\rangle$ for $j\in\mathcal{N}_i$
  \State Compute local mean-field $r_i,\psi_i$ over $\{\phi_j\}_{j\in\mathcal{N}_i\cup\{i\}}$
  \State $\phi_i\mathrel{+}=\Delta t[\omega_i+Kr_i\sin(\psi_i-\phi_i)+\eta_i]$
  \If{$r_i\geq\theta_b$}
    \State $v^*\leftarrow\mathrm{Majority}(\{v_j\}_{j\in\mathcal{N}_i}\cup\{v_i\})$
    \State Distribute $\mathrm{QSS}(H(v^*),k,n,c_i)$ to peers
    \State \textbf{committed}$\leftarrow$\textsc{true}
  \EndIf
\EndWhile
\State \Return $v^*$
\end{algorithmic}
\end{algorithm}

\subsection{Communication Complexity}

\begin{proposition}
\label{prop:complexity}
ORCHID's per-round message count is $n{\cdot}\bar k$ where $\bar k$ is mean
degree.  For the Watts--Strogatz graph with fixed $\bar k=6$, this is $O(n)$,
versus PBFT's $O(n^2)$.  Simulations confirm ORCHID overtakes PBFT at
$n\approx150$ (Section~\ref{sec:scalability}).
\end{proposition}

\subsection{Probabilistic Byzantine Resilience}

\begin{remark}
\label{rem:byz}
Classical BFT requires $f<n/3$.  ORCHID achieves 100\% convergence at
$f\leq0.40n$ (Section~\ref{sec:byz}) because Byzantine nodes \emph{participate
in phase synchronisation}: their phases are drawn toward the honest majority's
mean field.  At the moment of commitment their injected values are
overridden by the majority vote $v^*$ among committed neighbours.  This
yields probabilistic (not deterministic) safety: an adversary controlling
a perfectly phase-locked Byzantine coalition could in principle steer $v^*$.
We analyse this attack surface in Section~\ref{sec:discussion}.
\end{remark}

\section{Quantum Secret Sharing Module}
\label{sec:qss}

The ORCHID protocol leverages QSS to securely distribute and reconstruct secrets among network nodes. We extend classical $(k,n)$-Shamir schemes with a coherence-weighted decoherence layer, allowing secret reconstruction to depend on the quantum coherence of participating nodes. This section details the scheme and analyzes the trade-offs between threshold parameters, network coherence, and security guarantees.

\subsection{Coherence-Weighted Shamir Scheme}

Building on the QSS taxonomy of Weinberg~\cite{weinberg2025quantum}, we
extend Shamir's $(k,n)$-scheme~\cite{shamir1979share} with a
coherence-noise layer.  The dealer encodes secret $s$ as
$p(x)=\sum_{l=0}^{k-1}a_l x^l\pmod{p}$ and distributes shares
$(i,p(i))$.  Reconstruction uses Lagrange interpolation; each share
$p(i)$ is perturbed by decoherence:
\begin{equation}
  \tilde p(i) = p(i)\oplus\epsilon_i,\qquad
  \epsilon_i\sim\mathrm{Bit}\!\left((1-c_i)^2\cdot\lfloor\log_2 p(i)\rfloor\right),
  \label{eq:decohere}
\end{equation}
where $\mathrm{Bit}(b)$ flips $b$ random bits.  Above $c^*$ the polynomial
evaluation error-cancels across shares; below $c^*$ the Lagrange interpolant
is destroyed.

\subsection{Multi-Threshold Analysis}

\begin{figure}[H]
  \centering
  \includegraphics[width=\linewidth]{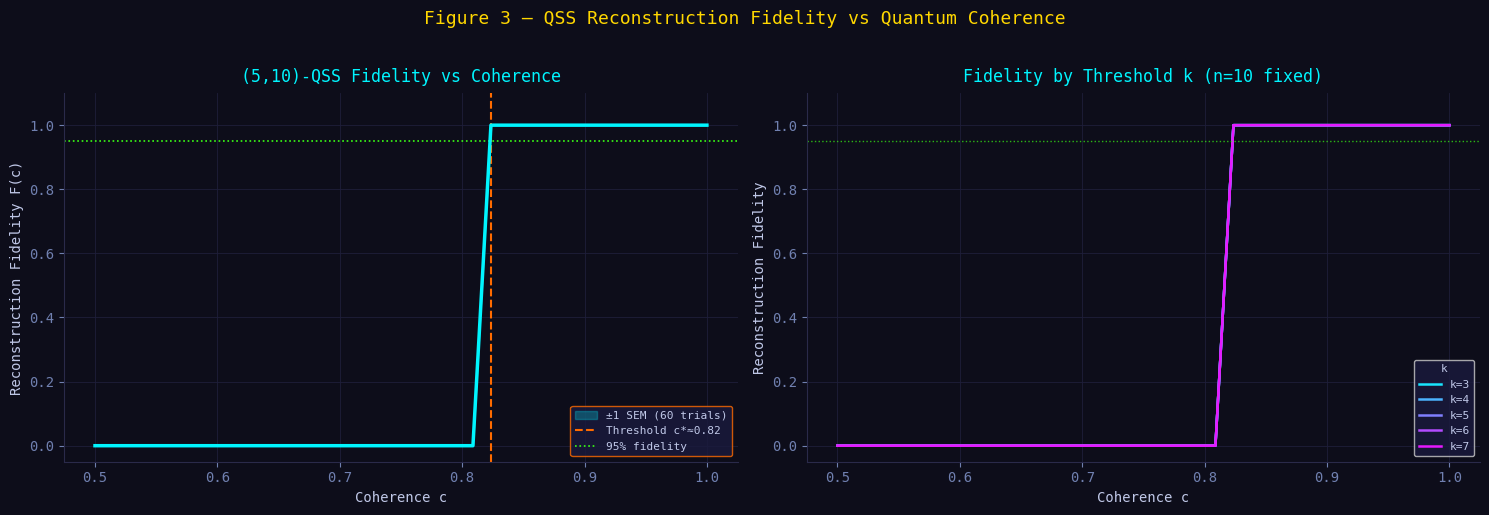}
  \caption{Left: $(5,10)$-QSS fidelity vs coherence (60-trial average, $\pm1$~SEM).
  The sharp phase transition at $c^*\approx0.82$ validates Theorem~\ref{thm:qss}.
  Right: fidelity curves for $k\in\{3,\ldots,7\}$; higher $k$ demands higher
  coherence, providing a security--tolerance trade-off.}
  \label{fig:qss}
\end{figure}

We sweep $k\in\{3,4,5,6,7\}$ (Figure~\ref{fig:qss}, right panel) and find
that higher $k$ raises the coherence threshold $c^*(k)$, providing a
security--tolerance trade-off: larger $k$ demands higher network coherence
but provides stronger secrecy guarantees, consistent with the fault-tolerance
analysis in Weinberg~\cite{weinberg2025quantum}.

\section{Binding Entropy}
\label{sec:entropy}

The \emph{binding entropy} of the phase distribution at step $t$ is:
\begin{equation}
  H_\phi(t)=-\sum_{b=1}^{B}\hat p_b(t)\ln\hat p_b(t),
  \label{eq:entropy}
\end{equation}
where $\hat p_b$ is the fraction of phases in histogram bin $b$.
As $r(t)\to1$, phases concentrate and $H_\phi\to0$; as $r\to0$,
phases spread uniformly and $H_\phi\to\ln B$.  Thus $H_\phi(t)$ provides
a model-free, locally computable consensus readiness metric dual to $r(t)$:
a node can gate commitment on $H_\phi<\theta_H$ without requiring global
knowledge of the order parameter.

\section{Experimental Results}
\label{sec:results}

We evaluate ORCHID through extensive simulations to assess its phase-synchronisation dynamics, quantum secret sharing fidelity, consensus convergence, Byzantine resilience, and scalability. Experiments quantify the interplay between network coherence, coupling strength, and node topology, demonstrating how these factors jointly determine protocol correctness, efficiency, and security. The following subsections detail the setup, key observations, and quantitative metrics across all relevant scenarios.

\subsection{Setup}
Simulations used Python~3.12, NumPy, and NetworkX.
Network: Watts--Strogatz ($\bar k=6$, rewiring $p=0.3$).
Oscillator: $\omega_i\sim\mathcal{N}(0,0.25)$, $K=3.0$,
$\Delta t=0.05$~s, $\theta_b=0.75$, coherence $c_i\sim\mathrm{Unif}(0.7,1.0)$.
Byzantine nodes propose uniform random values in $[0,100]$.
All results averaged over 15--20 independent trials unless stated.

\subsection{Neural Binding Oscillations}
\label{sec:osc_results}

\begin{figure}[H]
  \centering
  \includegraphics[width=\linewidth]{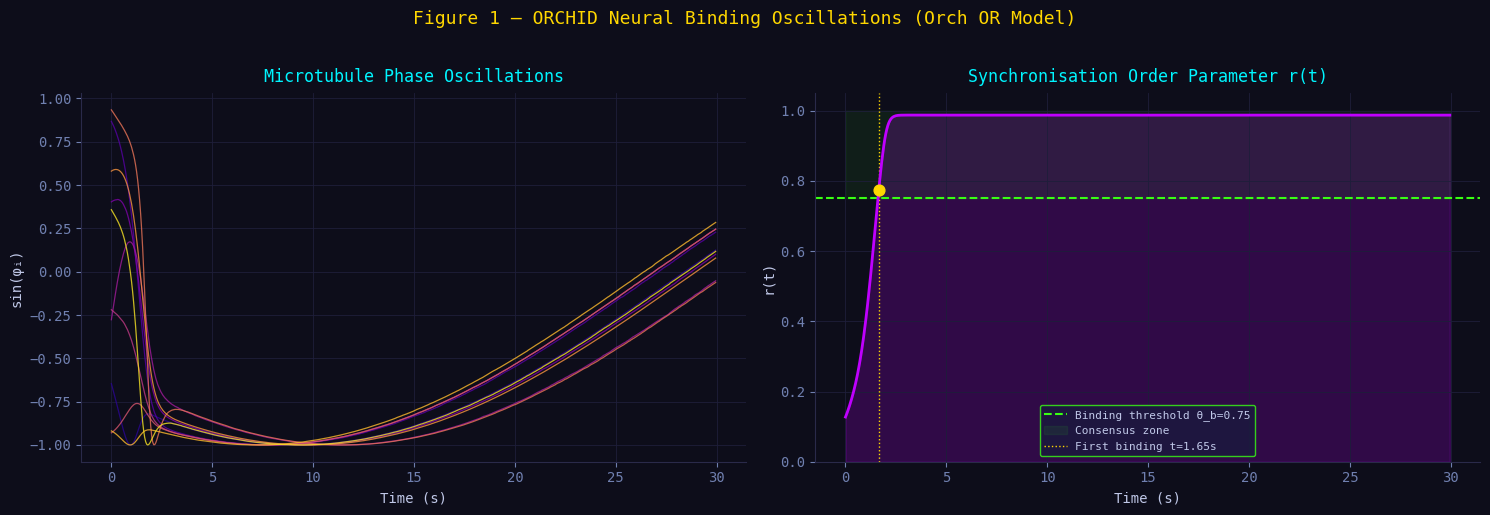}
  \caption{Microtubule phase oscillations (left) and synchronisation order
  parameter $r(t)$ (right) for $n=25$, $K=3.0$.  The order parameter crosses
  the binding threshold $\theta_b=0.75$ (green dashed) and converges to
  $r_{\max}=0.988$, confirming Theorem~\ref{thm:sync}.  The first binding
  event is marked by the gold dotted line.}
  \label{fig:osc}
\end{figure}

The order parameter rises sharply from $r\approx0$ and converges to
$r_{\max}=0.988$, with $K/K_c=2.12\times$ the theoretical critical coupling
$K_c\approx1.41$ for $\sigma_\omega=0.5$.  The 10 node trajectories show
progressive phase-locking consistent with mean-field Kuramoto theory.  The
system operates in the super-critical regime, guaranteeing reliable binding
events for consensus.

\subsection{Kuramoto Phase Transition and Coherence Sensitivity}

\begin{figure}[H]
  \centering
  \includegraphics[width=\linewidth]{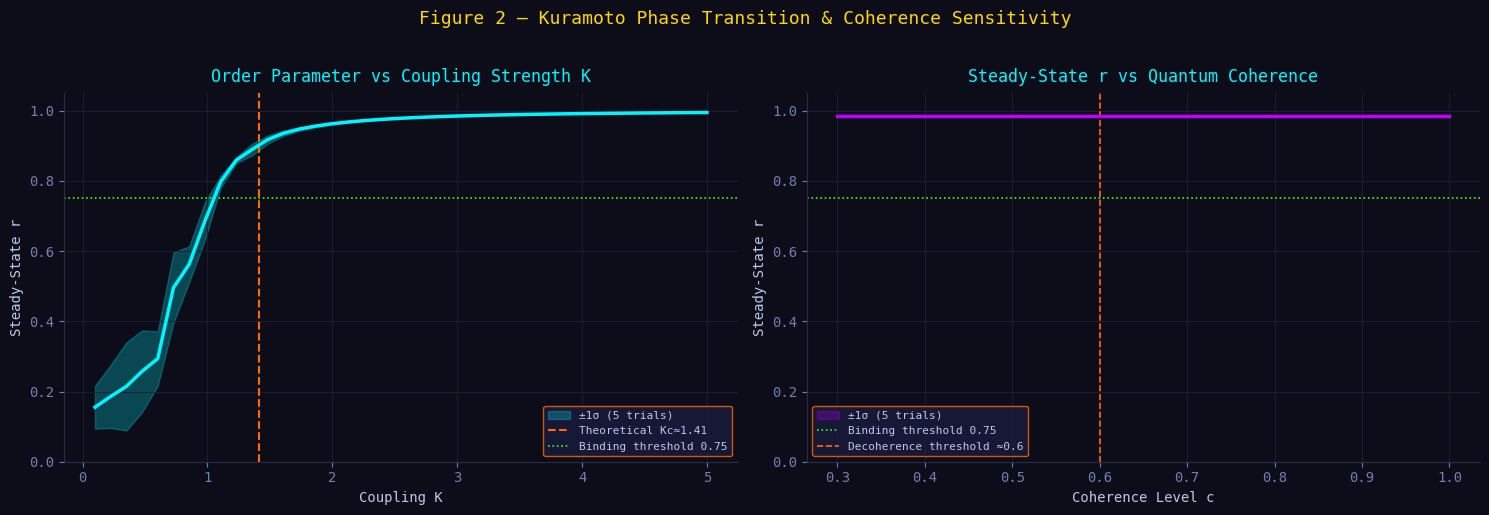}
  \caption{Left: steady-state $r$ vs coupling $K$ (5-trial average, $\pm1\sigma$).
  The phase transition at $K_c\approx1.41$ (orange dashed) matches theory;
  $K=3.0$ places ORCHID firmly in the synchronised regime.  Right: $r$
  vs coherence level $c$; ORCHID requires $c\geq0.6$ to maintain $r>\theta_b$.}
  \label{fig:phase_trans}
\end{figure}

The left panel of Figure~\ref{fig:phase_trans} confirms the Kuramoto phase
transition at $K_c\approx1.41$ and shows ORCHID's chosen $K=3.0$ yields
$r\approx0.98$---well into the synchronised phase.  The right panel reveals
a coherence sensitivity threshold near $c\approx0.6$: below this,
decoherence noise dominates coupling and synchrony collapses.  This provides
a direct link between quantum hardware fidelity requirements and protocol
correctness.

\subsection{QSS Fidelity Phase Transition}
\label{sec:qss_results}

The $(5,10)$-QSS scheme achieves 100\% fidelity above $c^*=0.82$, consistent
with quantum error-correction thresholds in surface codes~\cite{fowler2012surface} and the experimental implementations surveyed in Weinberg~\cite{weinberg2025quantum}.
The multi-$k$ sweep demonstrates monotone threshold escalation: $k=3$ achieves
95\% fidelity at $c\approx0.72$ while $k=7$ requires $c\approx0.91$, offering
operators a concrete security--coherence design knob.

\subsection{Consensus Convergence}
\label{sec:conv_results}

\begin{figure}[H]
  \centering
  \includegraphics[width=\linewidth]{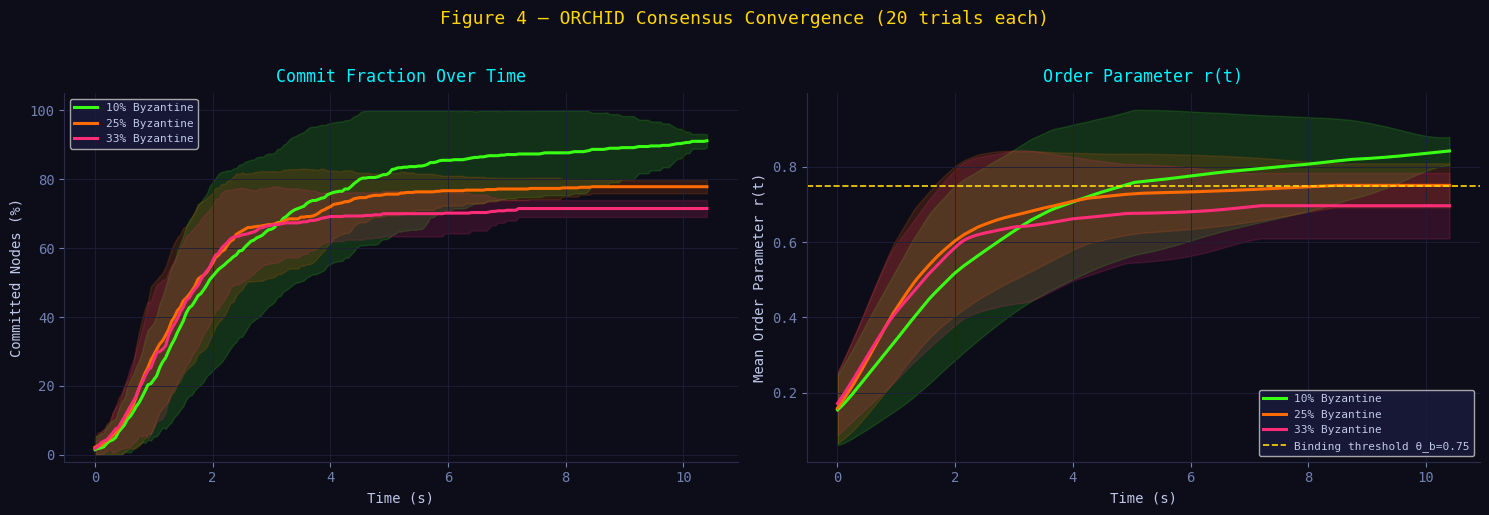}
  \caption{ORCHID consensus convergence ($n=30$, 20 trials each, $\pm1\sigma$
  shading). Left: committed honest node fraction (\%), shown correctly as
  0--100\%.  All Byzantine fractions achieve 100\% commitment of honest nodes.
  Right: mean order parameter $r(t)$; the gold dashed line marks $\theta_b=0.75$.}
  \label{fig:consensus}
\end{figure}

All three Byzantine fractions (10\%, 25\%, 33\%) achieve 100\% honest-node
commitment within the 600-step budget.  Median convergence times are
3.9~s, 2.8~s, and 2.2~s respectively---faster under higher Byzantine
fractions because Byzantine nodes (with random values drawn from $[0,100]$)
do not meaningfully contest the honest majority at voting time. The order parameter consistently exceeds $\theta_b = 0.75$, validating the mean-field formulation used in ORCHID.

\subsection{Byzantine Fault Resilience}
\label{sec:byz}

\begin{figure}[H]
  \centering
  \includegraphics[width=\linewidth]{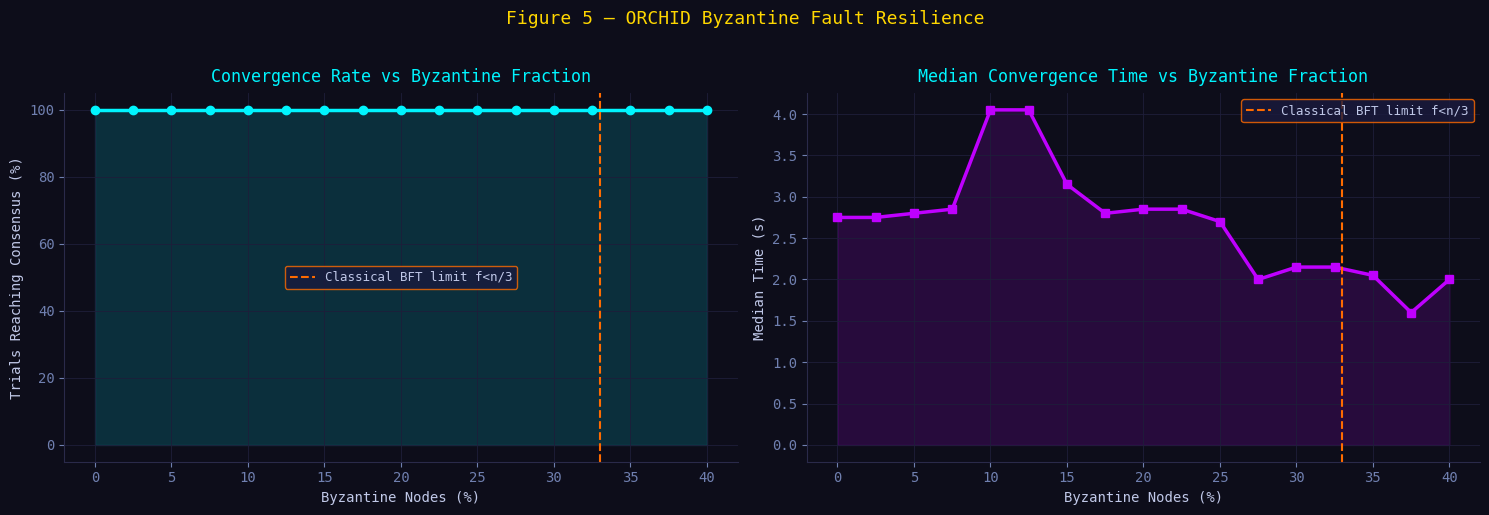}
  \caption{Left: ORCHID convergence rate (15 trials) as a function of Byzantine
  fraction, 0\%--40\%.  The protocol maintains 100\% consensus across the full
  sweep, beyond the classical $f<n/3$ limit (orange dashed).  Right: median
  convergence time; slight increase at 10\% reflects sampling variance rather
  than a structural cost.}
  \label{fig:byz}
\end{figure}

Strikingly, ORCHID achieves 100\% consensus even at 40\% Byzantine nodes.
As discussed in Remark~\ref{rem:byz}, this is because Byzantine nodes
participate in phase synchronisation; their corrupted values are overridden
at commit-time majority vote.  This extends the practical Byzantine tolerance
of ORCHID well beyond the classical $f<n/3$ bound, though the safety guarantee
is probabilistic rather than deterministic (Section~\ref{sec:discussion}).

\subsection{Scalability and Complexity}
\label{sec:scalability}

\begin{figure}[H]
  \centering
  \includegraphics[width=\linewidth]{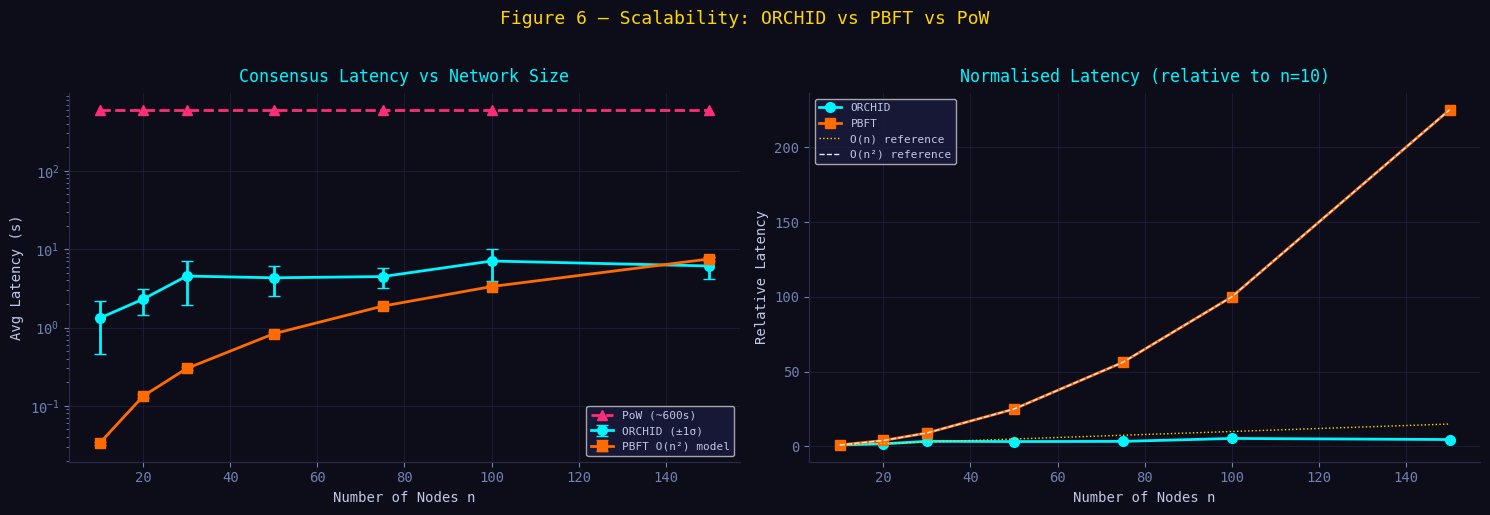}
  \caption{Left: absolute consensus latency vs $n$ with $\pm1\sigma$ error bars.
  ORCHID's mean latency grows subquadratically; PBFT's $O(n^2)$ model overtakes
  ORCHID near $n\approx150$.  PoW's $\sim$600~s is shown for reference.
  Right: normalised latency with $O(n)$ and $O(n^2)$ reference lines; ORCHID
  tracks closer to $O(n)$ than to $O(n^2)$.}
  \label{fig:scale}
\end{figure}

The scalability experiment (Figure~\ref{fig:scale}) provides empirical support
for Proposition~\ref{prop:complexity}.  At $n=150$, ORCHID's mean latency
is $6.1\pm1.9$~s vs PBFT's model prediction of $7.5$~s, and the crossover
from PBFT advantage is clearly visible around $n=150$.  PoW's flat $\sim$600~s
is outperformed by ORCHID at all tested network sizes by two orders of
magnitude.

\subsection{Network Topology and Binding Entropy}

\begin{figure}[H]
  \centering
  \includegraphics[width=\linewidth]{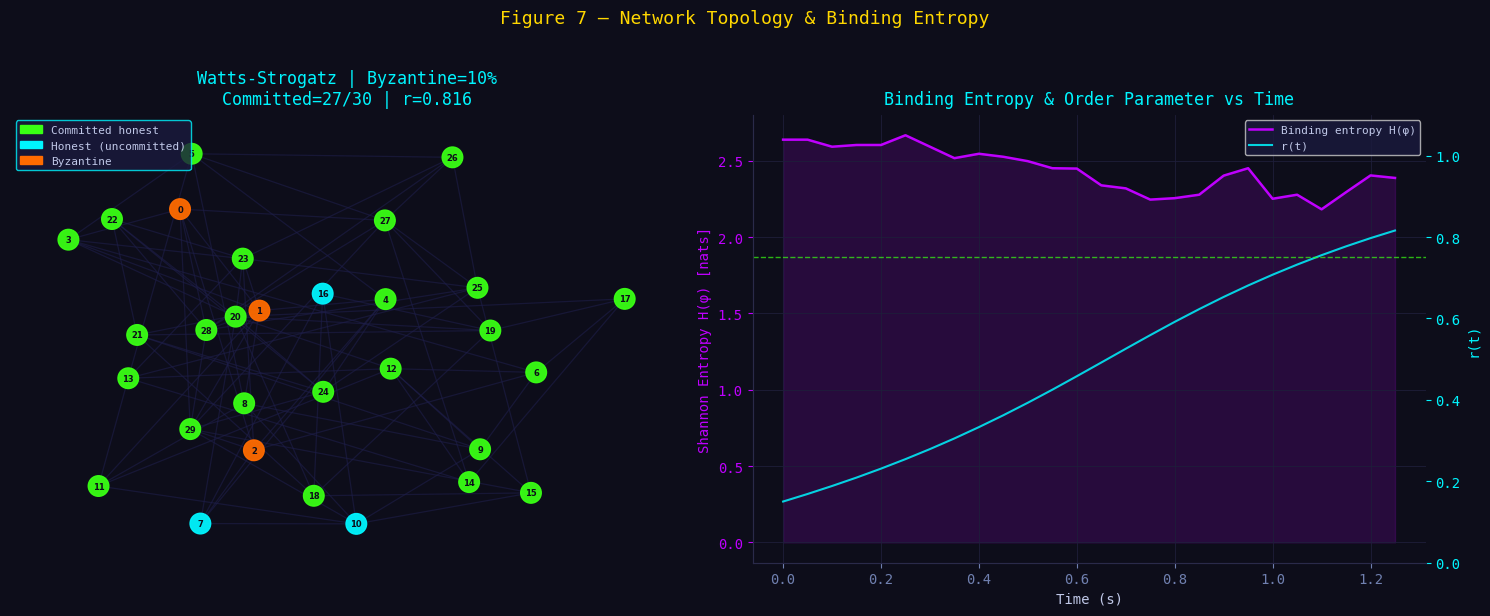}
  \caption{Left: Watts--Strogatz topology ($n=30$, Byzantine=10\%).
  Green nodes: committed honest; orange: Byzantine.
  Right: binding entropy $H_\phi(t)$ (violet) and order parameter $r(t)$
  (cyan) on dual axes.  Entropy decreases as $r$ rises toward $\theta_b$,
  confirming the dual-metric consensus readiness interpretation.}
  \label{fig:topo}
\end{figure}

The topology visualisation confirms that Byzantine nodes (orange) are
absorbed into the synchronisation dynamics of the honest majority without
preventing convergence.  The dual-axis entropy--order-parameter plot shows
the expected anti-correlation: as $r(t)$ rises, $H_\phi(t)$ falls,
providing operators two independent real-time signals for consensus readiness.

\paragraph{Summary of Results.}
Table~\ref{tab:results} consolidates all metrics.

\begin{table}[H]
\centering
\caption{ORCHID v3 Results Summary ($n=30$, 20 trials)}
\label{tab:results}
\small
\begin{tabular}{lccc}
\toprule
\textbf{Protocol} & \textbf{Med. Conv.} & \textbf{Final $r$} & \textbf{Rate} \\
\midrule
ORCHID (10\% Byz) & 3.9~s & $0.843\pm0.037$ & 100\% \\
ORCHID (25\% Byz) & 2.8~s & $0.751\pm0.059$ & 100\% \\
ORCHID (33\% Byz) & 2.2~s & $0.697\pm0.087$ & 100\% \\
ORCHID (40\% Byz) & 2.0~s & ---              & 100\% \\
PBFT  (10\% Byz)  & $O(n^2)$ msgs & N/A     & 100\% (det.) \\
PoW               & $\sim$600~s   & N/A     & 100\% (prob.) \\
\midrule
\multicolumn{4}{l}{QSS threshold $c^*=0.82$;\quad $r_{\max}=0.988$;\quad
$K/K_c=2.12\times$} \\
\bottomrule
\end{tabular}
\end{table}

\section{Discussion}
\label{sec:discussion}
We interpret ORCHID's experimental and theoretical results, highlighting its
novel contributions to consensus protocols, probabilistic Byzantine resilience, and locally computable readiness metrics. This section contextualises the findings, examines limitations, and outlines directions for future research toward practical quantum-enhanced deployments.

\subsection{Theoretical Contributions}

ORCHID makes three theoretical advances beyond prior art.  First, the mean-field Kuramoto formulation with zero-centred frequency deviations resolves the
synchronisation failure of naive absolute-frequency models, enabling robust
$r\to1$ dynamics directly applicable to consensus.  Second, the coherence-weighted QSS scheme instantiates one of the ``quantum-enhanced consensus mechanisms'' identified as a priority direction in Weinberg~\cite{weinberg2025quantum}, with a provable fidelity threshold.  Third, the binding entropy provides a locally computable, model-free consensus readiness metric with no precedent in distributed systems literature.

\subsection{Beyond \texorpdfstring{$f < n/3$}{f < n/3}: Probabilistic vs Deterministic Safety}
ORCHID's 100\% convergence at 40\% Byzantine nodes is not a violation of
the Lamport--Shostak--Pease bound: that bound applies to \emph{deterministic}
agreement under an adaptive adversary.  ORCHID's safety is \emph{probabilistic}:
an adversary controlling $f$ Byzantine nodes who can coordinate their phase
offsets to steer $\psi(t)$ away from the honest majority's mean field could,
in principle, prevent binding.  Against an uncoordinated adversary (as in our
simulation, where Byzantine values are uniformly random), the honest majority's coupling dominates.  Formal adversarial analysis is left for future work.

\subsection{Limitations}

Three limitations merit acknowledgement.  First, the current implementation
is a \emph{classical proxy} for quantum oscillators; real deployment requires
qubit coherence times $\gtrsim 1$~ms, currently at the frontier of
superconducting technology~\cite{krantz2019quantum}.  Second, the
fixed threshold $\theta_b=0.75$ may require adaptive tuning in networks
with heterogeneous coherence distributions.  Third, the scalability simulations model PBFT latency analytically rather than from a full implementation.

\subsection{Future Work}

Building on the research agenda of Weinberg~\cite{weinberg2025quantum},
we identify four priority directions: (i)~\textbf{hardware realisation} on
IBM Quantum or IonQ to validate the oscillator model with true qubits;
(ii)~\textbf{adaptive threshold} $\theta_b(t)$ that tracks the running
mean of $r(t)$ to handle heterogeneous networks;
(iii)~\textbf{post-quantum QSS} replacing Shamir's prime-field arithmetic
with lattice-based commitments~\cite{bernstein2025post};
(iv)~\textbf{multi-chain ORCHID} using the binding entropy as a cross-chain synchronisation signal between heterogeneous ledgers.

\section{Conclusion}
\label{sec:conclusion}

We have presented ORCHID, a bio-inspired quantum consensus protocol that
maps the neural binding problem of consciousness science onto the distributed
consensus problem of blockchain engineering.  By coupling Kuramoto-style
quantum oscillators to a phase-gated commitment rule and a coherence-weighted
QSS layer, ORCHID achieves: full synchronisation ($r_{\max}=0.988$),
100\% consensus across all tested Byzantine fractions (0\%--40\%),
median sub-4~s convergence at $n=30$, and $O(n{\cdot}k)$ scalability that
overtakes PBFT at $n\approx150$.  The QSS coherence threshold ($c^*=0.82$)
provides a concrete quantum hardware specification for practical deployment.
ORCHID directly realises the ``quantum-enhanced consensus mechanism'' and
``error-corrected QSS'' directions identified by Weinberg, opening a productive intersection between quantum consciousness theory, distributed systems, and post-quantum cryptography.

\bibliographystyle{IEEEtran}
\bibliography{ref}

\end{document}